\documentclass[twocolumn,preprintnumbers,amsmath,amssymb]{revtex4}
\usepackage{subfigure}
\usepackage{graphicx}% include figure files
\usepackage{dcolumn}% align table columns on decimal point
\usepackage{bm}% bold math

\begin{document}

\title{NETWORK STRUCTURE AND DYNAMICS, AND\\
EMERGENCE OF ROBUSTNESS BY STABILIZING SELECTION \\
IN AN ARTIFICIAL GENOME}

\author{Thimo Rohlf}
\affiliation{Santa Fe Institute, 1399 Hyde Park Road, Santa Fe, NM 87501, USA\\
Max-Planck-Institute for Mathematics in the Sciences, Inselstr. 22, D-04103 Leipzig, Germany}%
\author{Chris Winkler}%
% \email{chris.winkler@pioneer.com}
\affiliation{%
Pioneer Hi-Bred International, 7250 NW 62nd Ave., Johnston, IA 50131 USA
}%

\begin{abstract}
Genetic regulation is a key component in development, but a clear
understanding of the structure and dynamics of genetic networks is not
yet at hand. In this work we investigate these properties within an
artificial genome model originally introduced by Reil \cite{Reil1999}.
We analyze statistical properties of randomly generated genomes
both on the sequence- and network level, and 
show that this model correctly predicts the
frequency of genes in genomes as found in experimental data. 
Using an evolutionary algorithm based on stabilizing selection
for a phenotype, we show that robustness against single base mutations,
as well as against random changes in initial network states that mimic stochastic fluctuations in 
environmental conditions, can emerge in parallel. Evolved genomes exhibit
characteristic patterns on both sequence and network level.
\end{abstract}

\maketitle

\section{Introduction}
The transcription of DNA into mRNA and subsequent translation into
protein is the fundamental genetic process; it is the crucial first
step by which genetic information gives rise to an
organism. Development is not such a linear process, however. By
binding to specific regions of the genome, the protein produced by one
gene can affect the activity of other genes, and those genes may in
turn express proteins that enhance or inhibit still more genes. A
network of interactions responsible for the regulation of genetic
activity is thus defined. Such genetic regulation is important if
cells are to have independent control over their behavior.

Today, the available amount of data for regulatory interactions
in a number of model organisms, as, for example, Yeast \cite{Wagner2000}
is steadily increasing. A number of distinguishing
structural properties have been identified, namely scale-free
degree distributions \cite{Jordan2004}, motifs \cite{Dobrin2004} and modular organization \cite{Thieffry1999}.

Still, there is not enough information
to suggest a comprehensive theory of how genetic regulatory networks
attain a particular structure, how genes in such networks interact and
respond to perturbation, and how evolution has shaped these
factors. This study is an attempt to explore these questions in the
context of one particular model \cite{Reil1999}, in the hopes that it has
features that correspond to the limited data currently available, and
so that some progress toward a comprehensive theory might eventually
be made.

Traditionally, attempts to understand the characteristics of
regulatory networks have focused on dynamical properties. That is, a
network topology is specified and rules are applied to describe how
each gene in the network responds to inputs. Some initial state is
then assigned and the time evolution of gene activity is studied. A
variety of rules have been used, including Boolean switches
\cite{Kauffman1969}, thresholds \cite{Kurten1988b,Rohlf2002}, and differential equations
\cite{Glass1973}. Much less work has been done in understanding how the
machinery of transcription, translation, and binding might act
throughout the genome to produce the topology of a genetic network. In
fact, most studies of genetic networks ignore modeling DNA-specific
processes altogether \cite{Jong2002}. The first part of our study
examines to what extent Reil's model \cite{Reil1999}, which includes
explicit parameterizations for transcription and translation, can
produce realistic genetic networks based on random genome realizations.

A description of the method we will use for building genetic
regulatory networks follows, along with comparisons to published and
publically available experimental data. 
Statistical properties of random realizations of artificial genomes
are derived, and related to network structure.
Next, we investigate the
dynamics of our modeled networks when applying threshold dynamics to
gene behavior. Although this is a strong simplification, this type of
discrete dynamics has been successfully applied in a  number of studies that
are concerned with the co-evolution of network dynamics and -structure
\cite{Bornholdt1998,Bornholdt2000a,Ciliberti2007}. Finally, we are
interested in understanding the role evolution might play in selecting
particular network topologies. 
This is explored by asking how genome
structure changes when those networks with certain dynamical
properties are preferentially selected. 
Similar questions have been addressed in a small number of
previous proof-of-principle studies using artificial genomes\cite{Banzhaf2003a,Kuo2004a,Hallinan2004a}, however,
without relating the observed adaptation to changes in sequence
and network topology.
In particular, we investigate a
scenario of stabilizing selection similar to previous studies concerned
with the evolution of developmental canalization \cite{Ciliberti2007}. We find evolution
towards robustness of regulatory dynamics against both noise, modeled as fluctuating initial conditions,
and against mutations. We show that, in principle, this phenotypic robustness
can be traced back to adaptive changes on the sequence level
that lead to emergence of more robust regulatory networks. 

\begin{figure}[tb]
\includegraphics[width=8.5cm]{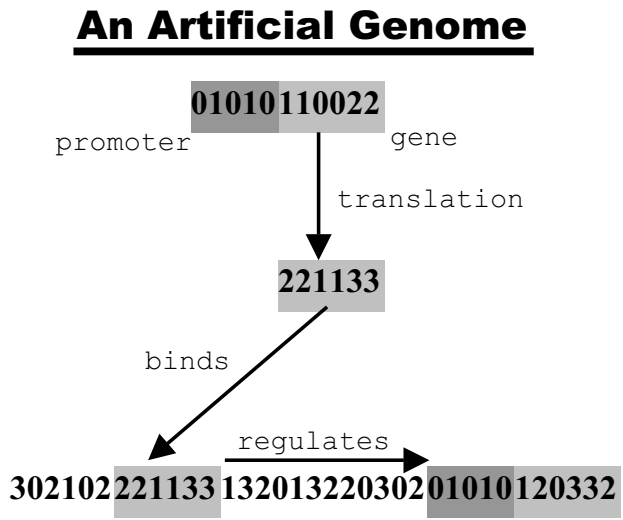}
\caption{\label{fig:schematic}Transcription, translation, and binding
in an artificial genome. The base promoter sequence is '01010' and is
indicated by dark shading. Light shading shows genes and
proteins. After \cite{Reil1999}.}
\end{figure}

\section{Model Details}

\subsection{Regulatory network construction from random sequences}
An artificial genome can be constructed as follows (also see
Fig.~\ref{fig:schematic}). Randomly string together $S$ integers drawn
uniformly between 0 and 3. The use of 4 digits need not be the case,
but does provide correspondence with the ATGC alphabet of real
genomes. 
%------------ Addition by TR, 28/10/2002-----------------
For the purpose of generalization, the length of the alphabet
in the artificial genome may in principle take any positive integer value $\lambda$.
%----------------------------------------------------------------
Next, define a base promoter sequence of length $l_p$ to
indicate the position of genes in the genome, say '01010'. Wherever
the promoter sequence occurs, the next $l_{g}$ digits are specified as
a gene. Translation of the gene sequence into a protein occurs
simply. Each number in the sequence is incremented by 1 and any values
greater than the last number in the base set of digits become the
first number ({\it e.g.}, the gene '012323' becomes the protein
'123030'). Binding sites are determined by searching the genome for
the protein sequence. If a match is found, then the protein is a
transcription factor (TF) that binds to that site and that regulates
the next downstream gene. TFs may enhance or inhibit gene activity. In
this study each TF has {\it equal} contribution to a gene's state and
has {\it equal} probability of activating or suppressing gene
expression. In real genetic systems, a TF may activate some genes
and inhibit others, depending on a complex interplay between various
factors that do not only depend on sequence. In our study,
we make the simplifying assumption that a TF is either activating
or inhibiting, which is determined by the sum $s_g$ of its sequence:
if $s_g < (1/2)s_{max}$, where $s_{max}$ is the maximal possible
cross sum value, it is inhibiting, otherwise it is activating.

Clearly this model greatly simplifies the true transcription,
translation, and binding processes. The binding of a transcription
factor to a cis-site, for example, depends on the protein's structure,
shape, and environment, rather than a simple template matching
approach. Moreover, there is a stochastic element to all these
processes that is simply ignored here.

Although it represents a strong simplification, the model does have biological
justification \cite{Reil1999}. The use of a base promoter sequence is
reminiscent of the TATA box frequently found in eukaryotic
organisms. Binding is modeled in a DNA-specific way, just as in real
organisms. Additionally, the model has the potential for greater
extendability than some models ({\it e.g.}, Boolean networks) because
it includes DNA-specific transcription, translation, and binding. The
impact of single base pair mutations on gene function and network
structure can be studied with this model, and also the effect
of sequence duplications (resulting in gene duplication) or -deletions
\cite{Leier2007}. In this paper, we will restrict ourselves
to single base pair mutations, and keep the genome size constant.

\subsubsection{Regulatory dynamics}
Dynamics of state changes (activity or inhibition of genes)
on the constructed networks can be defined in various ways. In our
study, we apply random threshold network (RTN) dynamics:
An RTN consists 
of $N$ randomly interconnected binary sites (spins) with states $\sigma_i=\pm1$.
For each site $i$, its state at time $t+1$ is a function of the inputs it receives 
from other spins at time $t$:
\begin{eqnarray} 
\sigma_i(t+1) = \mbox{sgn}\left(f_i(t)\right) 
\end{eqnarray}  
with 
\begin{eqnarray} 
f_i(t) = \sum_{j=1}^N c_{ij}\sigma_j(t) + h.  
\end{eqnarray}
The $N$ network sites are updated synchronously. In the following
discussion the threshold parameter $h$ is set to zero. The interaction weights
$c_{ij}$ take discrete values $c_{ij} = +1$ or $-1$ with equal
probability. If $i$ does not receive signals from $j$, one has $c_{ij} = 0$.

For a finite system size
$N$, the dynamics of RTN, which are closely related to Boolean
networks \cite{Kauffman1969} converge to periodic attractors (limit cycles)
after a finite number of updates. It has been suggested that different limit cycles
may correspond to different gene expression states (cell types) \cite{Kauffman1969}.
This property of RTN is also advantageous for defining phenotypes in
artificial evolutionary scenarios that are subject to various kinds of selective
pressure \cite{Ciliberti2007}.

\section{Statistical properties of the artificial genome}

In the following, $N$ denotes the number of genes in the artificial genome. 
$N$ is directly related to $S$, the number of bases, via the combinatorial
construction of the artificial genome.

\subsection{Genome size scaling}

Under the assumption that $l_g \ll \lambda ^{l_p}$, one derives easily the
following two relations:

\begin{equation} N(S,l_p) = \left(\frac{1}{\lambda}\right)^{l_p}\cdot S   \end{equation}
and
\begin{equation} S(N, l_p) =  \lambda ^{l_p}\cdot N\end{equation}

\subsection{Length of binding regions}
\subsubsection{Average length}
Constructing the artificial genome by random sampling from an alphabet of length $\lambda$,
the probability $p_p$ to get a promoter sequence after a given point of time is $p_p = \lambda^{-l_p}$.
Thus the expectation value for the length of the binding regions in Reil's artificial genome
is given by
\begin{equation}  \langle l_{bind} \rangle = \frac{1}{p_p} - l_g = \lambda^{l_p} -l_g.   \end{equation}

\subsubsection{Statistical distribution of $l_{bind}$}

To derive the exact statistical distribution of the lengths $l_{bind}$ of the binding regions
in the A.G., we first remark that the random production of promoter sequences during the
process of genome construction is a Bernoulli chain of length $n$ with the two possible
events: ``0'' - a certain base does not mark the begin of a promoter sequence and ``1'' - a certain 
base marks the start of a promoter sequence. Hence, the probability to produce $k$ promoters with $n$
random sampling steps is given by

\begin{equation} p(k,n) = {n \choose k} \cdot p_p^k(1-p_p)^{n-k}.  \label{pkneq} \end{equation}

By setting $k = 1$ and $n = l_{bind} + l_g + 1$ we get the probability $p_1$ to produce one promoter
within $n = l_{bind} + l_g + 1$ sampling steps:
\begin{equation} p_1 =  { l_{bind} + l_g + 1 \choose 1 }  \cdot p_p(1-p_p)^{l_{bind} + l_g }. \end{equation}
Only one of these possibilities is the correct one (production of a promoter with the \emph{last} sampling step),
i.e. for the derivation of $p(l_{bind})$ we have to divide Eq. (\ref{pkneq}) by the binomial coefficient, leading to

\begin{eqnarray} p(l_{bind}) & = & p_p(1-p_p)^{l_{bind} + l_g } \nonumber\\
                             & = & \lambda^{-l_p}(1-\lambda^{-l_p})^{l_{bind} + l_g} \\
                             & = &  \lambda^{-l_p}\exp{[-\alpha \cdot (l_{bind} + l_g )]},\end{eqnarray}
which is a decaying exponential distribution with $\alpha = -\ln{(1-\lambda^{-l_p})}$.

\begin{figure}[tb]
\centering
\subfigure[outdegree-distribution]{
\resizebox{60mm}{!}{\includegraphics{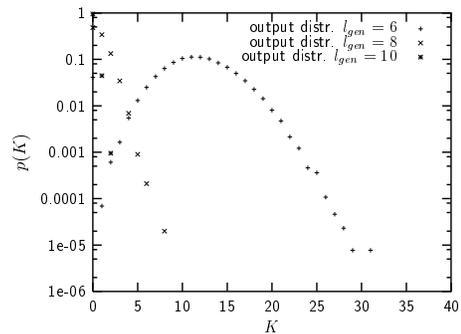}}
}
\subfigure[indegree
-distribution]{
\resizebox{60mm}{!}{\includegraphics{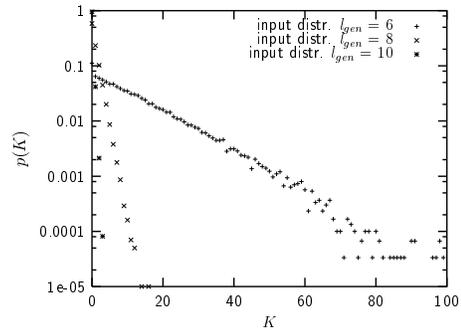}}
}
\caption{The probability of having $K_{out}$ regulatory outputs (a) and the probability of having $K_{in}$ regulatory inputs
for random genomes with different gene lengths $l_g$, averaged over $10^4$ realizations.}
\end{figure}

\subsection{Network connectivity}
In this section, we relate the previously derived properties of the artificial genome to characteristic parameters
of the resulting networks.

\subsubsection{Average connectivity}
For a given TF, the probability to match to a random ``DNA'' sequence of length $l_g$ is given by $p_{bind} = \lambda^{-l_g}$.
There are $\langle l_{bind}\rangle-l_g+1$ possibilities to bind to a binding region of average length $\langle l_{bind}\rangle$, 
thus the probability that
the TF provides at least one input to the gene defined by the promoter sequence following a given binding
region on average is 
\begin{eqnarray} \langle p_{input}\rangle &=& (\langle l_{bind}\rangle - l_g + 1)\cdot \lambda^{-l_g}\nonumber\\
                                          &=&  (\lambda^{l_p} - 2l_g + 1)\cdot \lambda^{-l_g}.\end{eqnarray}
Since we have $N$ binding regions, the average connectivity 
(averaged over the whole ensemble of possible random genomes) scales linear with the number of genes,
\begin{equation} \langle k\rangle = \langle p_{input}\rangle\cdot N,  \end{equation}
and the slope depends on $\lambda$, $l_g$ and $l_p$. 

Notice, however, that the average connectivity $\bar{K}$ obtained from a {\em particular} genome realization 
can substantially deviate from this typical value, since the possible values of $\bar{K}$ are Gaussian distributed around 
$\langle k\rangle$.

\subsubsection{In- and outdegree distribution}
From the above considerations, it is straight-forward to derive the statistical distributions for the number
of ingoing and outgoing links in randomly constructed genomes. 
Since a TF has equal {\em a priori} probability to bind at any region of the base string, generation of out-links is
a Poisson process, and hence the outdegree distribution is a Poissonian (Fig. 2a):
\begin{equation}
P(k_{out}) = \frac{\langle K\rangle ^{k_{out}}}{k_{out}!}\exp{[-\langle K\rangle]}.
\end{equation}
The number of inputs a gene receives, however, is proportional to the length $l_{bind}$ of its associated binding region,
hence, it follows from Eq.
\begin{equation}
P(k_{in}) \sim \exp{[-\beta k_{in}]} ,
\end{equation}
i.e. the indegree distribution is exponential. 
Both results are confirmed by numerical simulations (Fig. 2a  and 2b).

\begin{figure}[tb]
\includegraphics[width=8.5cm]{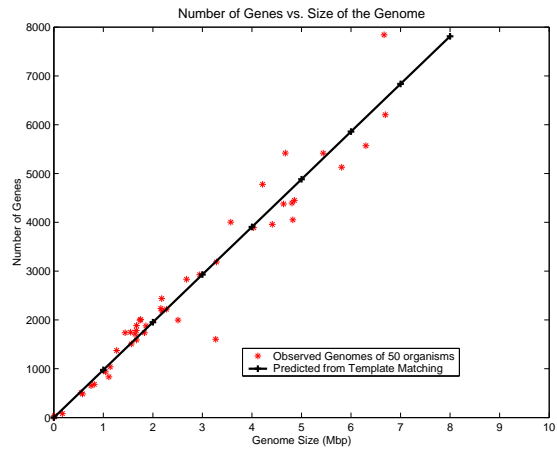}
\caption{\label{fig:numgenes}The number of genes predicted from the model as a function of
genome size $S$ with $l_{p}=5$. The number of genes in 50 organisms is
plotted for comparison. Observed data are taken from
http://www.ultranet.com/\~{}jkimball/.}
\end{figure}

\subsection{Relevance to biology}
Clearly, random genome realizations ar far from being a realistic model of real biological
genetic systems. However, it can be shown that even this extreme
oversimplification has some relevance for biology.  In Figure ~\ref{fig:numgenes}, the
predicted number of genes in a genome, $N=(1/4)^{l_p} \cdot S$, is plotted as a
function of genome size for $l_p=5$. Observed data
from 50 organisms that have been completely sequenced are also
shown. The correspondence between model and data is excellent for
this range of $S$ and shows that a combinatorial method for
determining the number of genes in a genome is appropriate. For larger
$S$, $l_{p}=7$ is reasonable (not shown), but little observed data exists.
On the other hand, statistical distributions of regulatory inputs and outputs
do not match biological data particularly well; here, more
realistic statistics can be obtained by constructing artificial genomes
from duplication and divergence events \cite{Leier2007}. However, even in these
models the question how selection pressure on the {\em phenotype},
as encoded by network dynamics, may influence genome organization,
remains unanswered. This type of question shall be addressed in the remaining part
of this paper.

\section{Stabilizing selection for a phenotype - an evolutionary scenario}
Though evolved by the random processes of genetic drift and selection pressure from changing environments, real genetic
systems are far from being random. Common wisdom is that this is often due to the highly non-linear
nature of the genotype-phenotype map, which includes an intermediate layer of complex regulatory processes
controlling cell machinery (unicellular organisms) or highly structured developmental processes (multicellular
organisms). The multilevel-structure of the involved evolutionary processes is sketched schematically in Fig. 4.
Typically, models of evolutionary adaptation focus either on sequence evolution or network structure alone,
and hence imply a huge loss of information as compared to the true multi-level and multi-scale evolutionary
dynamics.
Artifical genomes could be an important step towards models that integrate these levels, and hence may lead
to predictions on the effects of adaptive processes on sequence- and network evolution, and how these are related
to each other. 
In this section, we briefly explore an example of an evolutionary scenario based on an artificial genome,
motivated by the observation that development is highly canalized, i.e. buffered against both intrinsic and environmental
noise, and mutations \cite{Siegal2002}. A number of studies has demonstrated that stabilizing selection for particular phenotypes
leads to emergence of this high {\em robustness}, strongly fascilitated by the high amount of neutrality contained
in the fitness landscapes of complex regulatory networks \cite{Fernandez2007}. Let us now define an evolutionary algorithm of stabilizing selection
in a strongly fluctuating environment, based on an artificial genome.
We start by generating an initial population of randomly assembled genomes; the number of bases
is constrained such that each string contains exactly 64 genes. Next, different limit cycles of the
associated RTNs are identified by running network dynamics, as defined in section 2.1.1, from 10000 different
random initial state configurations. This process is stopped when a RTN is identified which has a fixed point
$S_f$ (a limit cycle  of length one), and at least 5 additional attractors; the relative weight of the basin of attraction
leading to $S_f$ should be small (less than $40\%$ of the tested configurations). The last two criteria are chosen
to rule out a to quick convergence of evolutionary dynamics (i.e., to make the problem hard). $S_f$ is the phenotype
we want to stabilize, and the digit string $G_f$, that codes for its regulatory network, is the genotype we evolve.

\begin{figure}[tb]
\includegraphics[width=8.5cm]{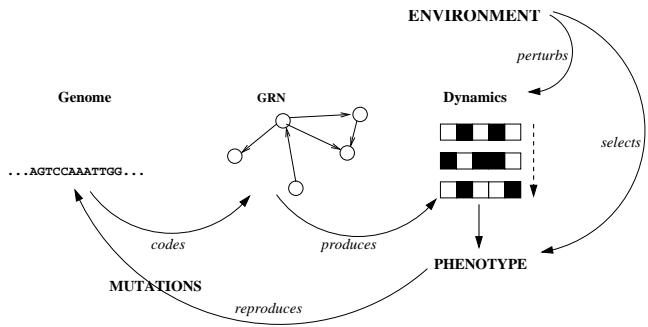}
\caption{Multilevel structure of the evolving genotype-phenotype map, involving an intermediate
level of complex regulatory networks.}
\end{figure}

\begin{figure}[tb]
\centering
\subfigure{
\resizebox{65mm}{!}{\includegraphics{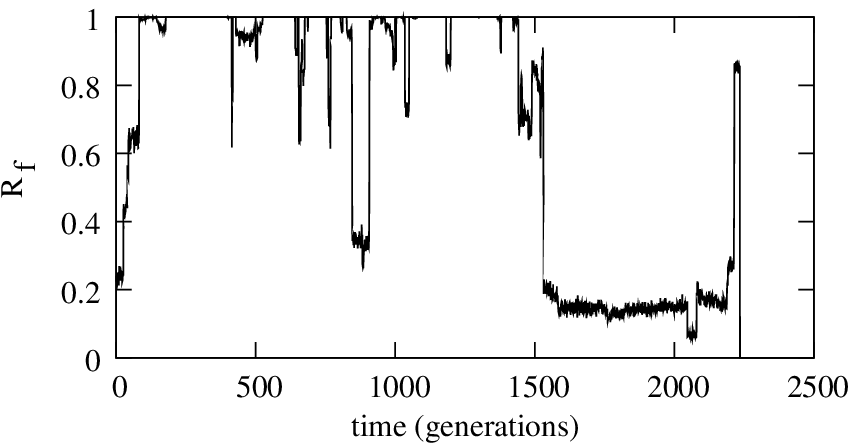}}
}
\subfigure{
\resizebox{65mm}{!}{\includegraphics{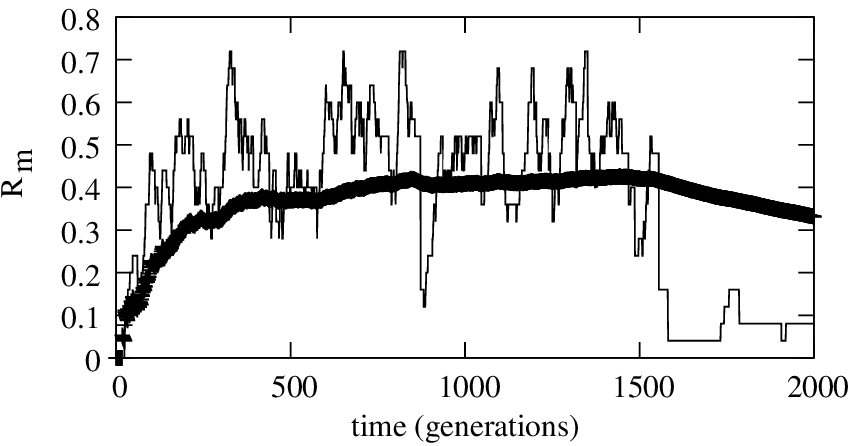}}
}
\caption{Example of an evolutionary run. Left: Evolution of the robustness $R_f$ against fluctuations
in initial conditions. Right: Evolution of the mutational robustness $R_m$ (thick line: cumulative average).}
\end{figure}

\begin{figure}[htb]
\includegraphics[width=8.5cm]{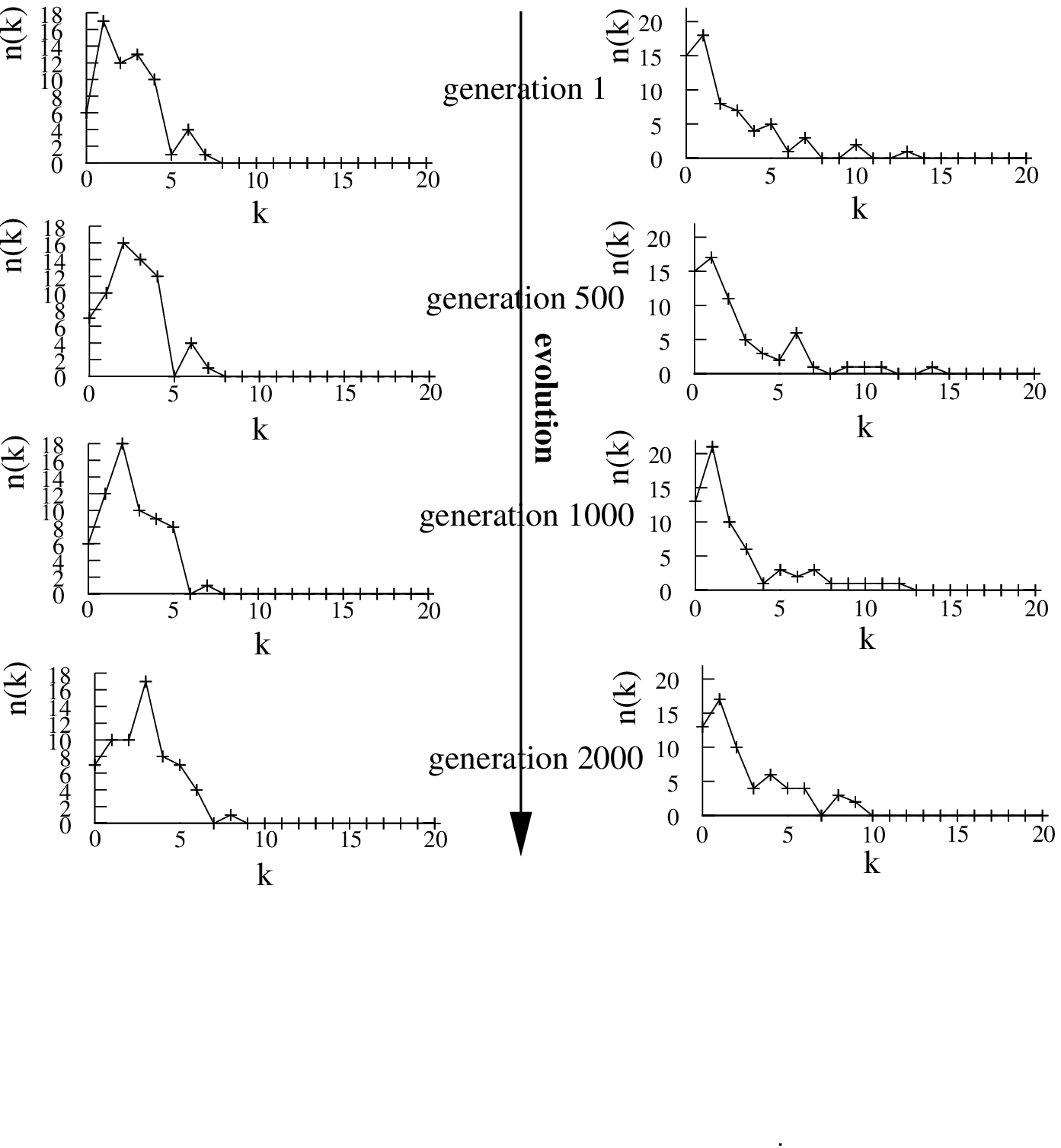}
\vspace*{-2cm}
\caption{Left: evolution of the outdegree-distribution, right: evolution of the indegree-distribution
in the same evolutionary run as in Fig. 5.}
\end{figure}

We now apply stabilizing selection as follows:
\begin{enumerate}
\item Create a mutant $\bar{G}_f$ by random single base mutations, occurring with a probability $p_m$ per base.
\item Run RTN dynamics from a random initial state, until an attractor is reached, otherwise stop after 200
iterations.
\item If dynamics has converged to $S_f$, keep $\bar{G}_f$, otherwise keep $G_f$.
\item For the next generation, iterate from (1).
\end{enumerate}
We note that we disregard mutations of promoter sites, as well as mutation leading to new "genes", to avoid
complications resulting from a varying genome size.
Notice that, in step (2), we test only one initial configurations, corresponding to the fact that
biological organisms are tested only against the environment they face at the {\em current} generation.
Robustness against fluctuations, i.e. the capacity to stabilize the phenotype in a large number of possible environments,
is measured by running RTN dynamics for $G_f$ ($\bar{G}_f$) for a larger set $Z$ of initial configurations (e.g. $10^4$ random initial states).
Then 
\begin{equation}
R_f(t) := \frac{Z_f(t)}{Z}
\end{equation}
defines the robustness against fluctuations, where $Z_f(t)$ is the fraction of initial states that lead to $S_f$
at generation $t$.
A second measure of robustness is associated to the capacitance to buffer the system against disadvantageous
mutations (mutational robustness $R_m$, \cite{Ciliberti2007}). At each generation we measure
the number of accepted mutants $P_a$ in the previous $P$ generations, and define
\begin{equation}
R_m(t) := \frac{P_a(t)}{P}.
\end{equation}

\begin{figure}[tb]
\includegraphics[width=8.5cm]{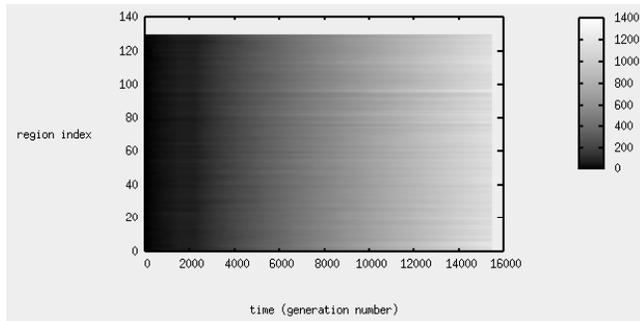}
\caption{Number of base exchanges during evolution for different positions on the genome,
averaged over regions containing 100 bases each. 
The brightness in grayscale indicates the number of bases exchanges. Shaded lines running from left to right
indicate conserved regions. }
\end{figure}

If $P_a$, and hence $R_m$ increases with $t$, this indicates restructuring of the genome such that
the probability of neutral or advantageous mutations with respect to $S_f$ has increased.
Fig. 5 shows both quantities in a typical evolutionary run. Both $R_f$ and $R_m$ increase rapidly,
however, exhibiting considerable fluctuations. In particular, $R_f$ exhibits an interesting
intermittent dynamics reminiscent of a {\em punctuated equilibrium} \cite{Bornholdt1998}, indicating
metastability of the evolutionary dynamics. In fact, in all evolutionary runs we studied $R_f$ and
$R_m$ could be stabilized only over a finite number of generations, as indicated in Fig. 5
by the sharp decrease of both quantities around $t = 1500$. $R_f$ and $R_m$ are positively
correlated, similar to the results reported in \cite{Ciliberti2007}. The evolutionary instability is an inevitable consequence
of the fact that $S_f$, in each generation, is tested only against a very limited set of mutations
and environments. The artificial genome now allows us to trace the effects of this non-trivial evolutionary dynamics on
both network and sequence structure.

Figure 6 shows the evolution of the distributions of regulatory input and output numbers per gene, in the same
evolutionary run as shown in Fig. 5 with regards to adaptation dynamics. Evidently, considerable
reorganization of network structure is taking place: while the indegree-distribution tends to become narrower,
the peak of the outdegree-distribution is shifted towards larger $k$. However, these trends are not particularly
pronounced, probably due to the small genome size applied ($N=64$ genes). Interestingly,
sequence information turns out to be more informative for generating hypotheses how robustness evolves.
Figure 7 shows the number of base exchanges during evolution for different positions on the genome.
At each generation, the cumulative number of base changes in successive slices of 100 digits on the genome string
during all previous generations was monitored. Different gray shades in Fig. 7, that are maintained
over the whole evolutionary run, indicate that there are conserved regions, while in other regions
base changes accumulate more rapidly. While we do not yet have a conclusive
explanation for this observation, a tentative hypothesis would be that the conserved regions encode binding sites
of genes that regulate the invariant "core dynamics" of the phenotype, while the regions with more frequent
base substitutions are responsible for stabilizing it against fluctuations, or contain
mostly neutral mutations. A detailed analysis of the correlations between
sequence- and network evolution, which goes beyond the scope of our present study, may shed more light on this problem.

%\section{Summary and Outlook}

\bibliography{agenome_bib1_140408}{}

\begin{thebibliography}{10}

\bibitem{Banzhaf2003a}
Wolfgang Banzhaf.
\newblock On the dynamics of an artificial regulatory network.
\newblock In W.~Banzhaf, T.~Christaller, P.~Dittrich, J.~Kim, and J.~Ziegler,
  editors, {\em Advances in Artificial Life, Proceedings of the 7th European
  Conference (ECAL-2003), Dortmund, September 15-17, 2003}, Lecture Notes in
  Artificial Intelligence, LNAI 2801, pages 217--227. Springer, Berlin, 2003.

\bibitem{Bornholdt2000a}
S.~Bornholdt and T.~Rohlf.
\newblock Topological evolution of dynamical networks: Global criticality from
  local dynamics.
\newblock {\em Phys. Rev. Lett.}, 84:6114--6117, 2000.

\bibitem{Bornholdt1998}
S.~Bornholdt and K.~Sneppen.
\newblock Neutral mutations and punctuated equilibrium in evolving genetic
  networks.
\newblock {\em Phys. Rev. Lett.}, 81:236--239, 1998.

\bibitem{Ciliberti2007}
Stefano Ciliberti, Oliver~C. Martin, and Andreas Wagner.
\newblock Robustness can evolve gradually in complex regulatory networks with
  varying topology.
\newblock {\em PLoS Computational Biology}, 3:e15, 2007.

\bibitem{Jong2002}
Hidde de~Jong.
\newblock Modeling and simulation of genetic regulatory systems: A literature
  review.
\newblock {\em J. Comp. Biol.}, 9:67--103, 2002.

\bibitem{Dobrin2004}
Radu Dobrin, Quasim~K. Beg, A.~L. Barab\'{a}si, and Z.~N. Oltvai.
\newblock Aggregation of topological motifs in the escherichia coli
  transcriptional regulatory network.
\newblock {\em BMC Bioinformatics}, 5:10, 2004.

\bibitem{Fernandez2007}
Pau Fernandez and Ricard Sole.
\newblock Neutral fitness landscapes in signalling networks.
\newblock {\em J. R. Soc. Interface}, 4:41--47, 2007.

\bibitem{Glass1973}
Leon Glass.
\newblock The logical analysis of continous, non-linear biochemical control
  networks.
\newblock {\em J. Theor. Biol.}, 39:103--129, 1973.

\bibitem{Hallinan2004a}
J.~Hallinan and J.~Wiles.
\newblock Asynchronous dynamics of an artificial genetic regulatory network.
\newblock In {\em Ninth International Conference on the Simulation and
  Synthesis of Living Systems (ALife9)}, September 2004.

\bibitem{Jordan2004}
I.K. Jordan, L.~Mariño-Ramírez, Y.I. Wolf, and E.V. Koonin.
\newblock Conservation and coevolution in the scale-free human gene
  coexpression network.
\newblock {\em Mol Biol Evol.}, 21:2058--2070, 2004.

\bibitem{Kauffman1969}
S.A. Kauffman.
\newblock Metabolic stability and epigenesis in randomly connected nets.
\newblock {\em J. Theor. Biol.}, 22:437--469, 1969.

\bibitem{Kuo2004a}
Paul~Dwight Kuo, Andre Leier, and Wolfgang Banzhaf.
\newblock Evolving dynamics in an artificial regulatory network model.
\newblock In Yao X., Burke E., Lozano J.A., Smith J., Merelo-Guervós J.J.,
  Bullinaria J.A., Rowe J., Tino P., Kabán A., and Schwefel H.-P., editors,
  {\em Proc. of the Parallel Problem Solving from Nature Conference (PPSN-04),
  Birmingham, UK, September 2004}, pages 571--580. Springer, LNCS 3242, Berlin,
  2004.

\bibitem{Kurten1988b}
K.E. K\"urten.
\newblock Correspondence between neural threshold networks and kauffman boolean
  cellular automata.
\newblock {\em J. Phys. A}, 21:L615--L619, 1988b.

\bibitem{Leier2007}
A.~Leier, D.P. Kuo, and W.~Banzhaf.
\newblock Analysis of preferential network motif generation in an artificial
  regulatory network model created by duplication and divergence.
\newblock {\em Advances in Complex Systems}, 10:155 -- 172, 2007.

\bibitem{Reil1999}
T.~Reil.
\newblock Dynamics of gene expression in an artificial genome - implications
  for biological and artificial ontogeny.
\newblock In {\em Proceedings of the 5th European Conference on Artificial
  Life}, pages 457--466. Springer, 1999.

\bibitem{Rohlf2002}
T.~Rohlf and S.~Bornholdt.
\newblock Criticality in random threshold networks: Annealed approximation and
  beyond.
\newblock {\em Physica A}, 310:245--259, 2002.

\bibitem{Siegal2002}
Mark~L. Siegal and Aviv Bergman.
\newblock Waddington's canalization revisited: Developmental stability and
  evolution.
\newblock {\em Proc. Natl. Acad. Sci.}, 99:10528--10532, 2002.

\bibitem{Thieffry1999}
D.~Thieffry and D.~Romero.
\newblock The modularity of biological regulatory networks.
\newblock {\em BioSystems}, 50:49--59, 1999.

\bibitem{Wagner2000}
A.~Wagner.
\newblock Robustness against mutations in genetic networks of yeast.
\newblock {\em Nature Genetics}, 24:355--361, 2000.

\end{thebibliography}
\bibliographystyle{plain}

\end{document}